\documentclass[12pt]{amsart}

\usepackage{amsmath,amssymb,amsthm,amsfonts}

\newtheorem{thm}{Theorem}
\newtheorem{prop}{Proposition}
\newtheorem{defin}{Definition}

\newtheorem{cor}{Corollary}
\newtheorem{lem}{Lemma}
\newtheorem{conj}{Conjecture}
\newtheorem{rem}{Remark}
\newtheorem{exa}{Example}

\newcommand{\B}{{\mathcal B}}

\newcommand{\C}{{\mathbb C}}

\newcommand{\F}{{\mathbb F}}

\newcommand{\HH}{{\mathbb H}}
\newcommand{\hh}{{\mathfrak H}}

\newcommand{\M}{{\mathcal M}}

\newcommand{\N}{{\mathbb N}}

\newcommand{\oo}{{\mathfrak O}}

\newcommand{\R}{{\mathbb R}}

\renewcommand{\u}{{\mathfrak U}}\newcommand{\U}{{\mathcal U}}

\newcommand{\W}{{\mathcal W}}

\newcommand{\doc}{\mathcal{D}\mathcal C}   
\newcommand{\diag}{\text{Diag}}   

\newcommand{\id}{\text{Id}}   
\newcommand{\iso}{\text{Iso}}   

\newcommand{\os}{\mathcal O\mathcal S}

\newcommand{\de}{\delta}

\newcommand{\la}{\lambda}

\newcommand{\tv}{\tilde{v}}

\date{\today}

\begin{document}

\title[Generalized unistochastic matrices]
{On a multi-dimesional generalization of the notion of
orthostochastic and unistochastic matrices}

\author{Eugene Gutkin}

\address{Nicolaus Copernicus University, Department of Mathematics, Chopina 12/18,
87-100 Torun and Mathematics Institute of the Polish Academy of
Sciences, Sniadeckich 8, 00-956 Warsaw, Poland}

\email{gutkin@mat.umk.pl,\ gutkin@impan.pl}

\date{\today}

\begin{abstract}
We introduce the notions of $d$-orthostochastic,
$d$-unistochastic, and $d$-qustochastic matrices. These are the
particular cases of $\F^d$-bistochastic matrices for
$\F=\R,\C,\HH$. The concept is motivated by mathematical physics.
When $d=1$, we recover the orthostochastic, unistochastic, and
qustochastic matrices respectively.  This work exposes the basic
properties of $\F^d$-bistochastic matrices.
\end{abstract}

\maketitle

\tableofcontents

\section{Introduction}  \label{intro}
A square matrix $P$ with nonnegative entries $p_i^j$ adding up to
one in every row and column is called a {\em bistochastic matrix}.
Bistochastic matrices come up in probability, combinatorics,
mathematical physics, geometry, optimization, etc
\cite{An82,HoJo,BZ06,Gu11}. Let $\F$ stand for $\R,\C$ or $\HH$,
let $n\in\N$, and let $\B_n$ (resp. $U(\F,n)$) denote the set of
$n\times n$ bistochastic matrices (resp. the group of linear
isometries of $\F^n$). The matrix of squared norms of entries of
$V\in U(\F,n)$ is bistochastic, yielding the {\em squared norm
mapping} $\nu:U(\F,n)\to\B_n$.

The ranges of $\nu$ for $\F=\R,\C,\HH$ are the sets of {\em
orthostochastic, unistochastic, qustochastic} $n\times n$
matrices, denoted by $\oo_n,\U_n,\hh_n$ respectively. The obvious
inclusions $\oo_n\subset\U_n\subset\hh_n\subset\B_n$ are proper if
$n\ge 4$ \cite{ChDo08}. There are connections between these
matrices and (generalized) numerical ranges and numerical shadows
\cite{Toeplitz,Hausdorff,GJK04,JAGut98,DGHPZ11,GS10,GuRa,Gu04,P+12,GuZy13,KRS97}.
They (especially unistochastic matrices) are of importance in
quantum physics \cite{BZ06,Betal05,P+12,DuZy11}. There is a
considerable literature regarding orthostochastic, unistochastic,
and qustochastic matrices {\cite{Na96,ChDo08,AuPo79,DuZy11}.
However, several basic questions remain open \cite{ChDo08,DuZy11}.

The motivation for the present work comes from mathematical
physics. The entries of a bistochastic matrix $P\in\B_n$ are the
probabilities of transition between the states of a classical
physical system. {\em Transitions amplitudes} in quantum physics
are complex numbers, and the {\em transition probabilities} are
their squared absolute values. The transitions amplitudes of a
quantum system with $n$ basic states form a unitary $n\times n$
matrix $U$. Let $P=\nu(U)$ be the matrix of squared norms of the
entries in $U$. Then $P$ is a bistochastic matrix describing the
corresponding classical system. Thus, a bistochastic matrix $P$ is
unistochastic if the classical system described by $P$ can be
quantized, yielding a quantum physical system whose transition
amplitudes are the entries of a unitary matrix $U$ satisfying
$\nu(U)=P$.

Suppose now that the basic states of a quantum system have
internal degrees of freedom. Assume, for simplicity, that the
number, say  $d>1$, of internal degrees of freedom is the same for
all states. The transition amplitudes become vectors
$v_i^j\in\C^d$. This quantum system corresponds to a $n\times n$
matrix $V=[v_i^j]$ with vector entries. The unitarity condition
says that the operator $V:\C^n\to\C^{nd}$ is an isometry. Let
$\iso(\C,n,d)$ denote the set of these isometries. For
$V\in\iso(\C,n,d)$ the corresponding bistochastic $n\times n$
matrix $P$ is given by $p_i^j=||v_i^j||^2$. Thus, our quantum
systems correspond to $V\in\iso(\C,n,d)$; the reduction from a
quantum to the classical system is given by the squared norm map
$\nu:\iso(\C,n,d)\to\B_n$. Its range is the set
$\os(\C,n,d)\subset\B_n$ of {\em  $d$-unistochastic matrices}. A
classical system described by $P\in\B_n$ admits a {\em
quantization with $d$ internal degrees of freedom} if and only if
$P\in\os(\C,n,d)$.

The same construction, based on $\R$ (resp. $\HH$) leads to {\em
$d$-orthostochastic matrices $\os(\R,n,d)$} (resp. {\em
$d$-qustochastic matrices $\os(\HH,n,d)$}). To streamline the
exposition, we will refer to $\os(\F,n,d)\subset\B_n$ as the set
of {\em $(\F,n,d)$-bistochastic} or simply {\em
$\F^d$-bistochastic matrices}.

We have structured the exposition as follows. In
section~~\ref{prelim} we precisely define the above notions and
expose a few basic properties of vector $\F^d$-bistochastic
matrices. Section~~\ref{general} continues this exposition.
Section~~\ref{vector_ortho} is devoted to $(n-1)$-orthostochasic
matrices. It contains our main results, Theorem~~\ref{d_min_thm}
and Corollary~~\ref{open_set_cor}, and a few examples. In the
concluding section~~\ref{conj} we state a conjecture regarding
$d$-orthostochastic matrices.

\medskip

\section{Definitions and basic properties} \label{prelim}
A $m\times n$ matrix is a collection of $mn$ entries organized in
$m$ rows and $n$ columns. In general, we will denote matrices by
capital letters, and denote their entries by the corresponding low
case letters with subscripts and superscripts. For instance, let
$P=[p_i^j]$ be a $n\times n$ matrix. The lower (resp. upper) index
stands for the column (resp. row). The entries of a matrix may be
numbers, vectors, etc.

We will denote by $\F$ any of the fields $\R$ or $\C$ or $\HH$.
Recall that a real $n\times n$ matrix $P=[p_i^j]$ is {\em
bistochastic} if $p_i^j\ge 0$ and $\sum_{j=1}^n p_i^j=\sum_{i=1}^n
p_i^j=1$. We will denote by $\B_n$ the set of bistochastic
$n\times n$ matrices.

For $N\in\N$ let $\F^N$ denote the vector space of $N$-tuples over
$\F$. Let $d,n\in\N$. The space $\F^{nd}$, decomposed as the
direct sum of $n$ copies of $\F^d$ will be denoted either by
$\oplus_{i=1}^n\F_i^d$ or simply by $\F^{nd}$ if the decomposition
is clear from the context.

Let $\M(\F^d,n)$ be the space of $n\times n$ matrices with entries
in $\F^d$.

\begin{lem}  \label{colum_row_lem}
A matrix $V=[v_i^j]\in\M(\F^d,n)$ satisfies  for all $1\le j,k\le
n$ the equation
\begin{equation}   \label{isom_eq}
\sum_{i=1}^n\langle v_j^i,v_k^i\rangle = \de(j,k).
\end{equation}
if and only if  for all $1\le j,k\le n$ we have
\begin{equation}   \label{conjug_eq}
\sum_{i=1}^n\langle v_i^j,v_i^k\rangle = \de(j,k).
\end{equation}
\begin{proof}
Let $\langle\,,\rangle_n$ and $\langle\,,\rangle_{nd}$ denote the
scalar products in $\F^n$ and $\F^{nd}$ respectively. Any
$V\in\M(\F^d,n)$ determines a linear operator $V:\F^n\to\F^{nd}$.
The adjoint operator $V^*:\F^{nd}\to\F^{n}$ satisfies for all
$x\in\F^{n},y\in\F^{nd}$ the identity
\begin{equation}      \label{adjoint_eq}
\langle Vx,y\rangle_{nd}=\langle\,x,V^*y\rangle_{n}.
\end{equation}
Let $\id_k$ denote the identity operator on $\F^k$.
Equation~~\eqref{isom_eq} means that $V:\F^n\to\F^{nd}$ is an
isometry, i.e.
$$
\langle Vx,Vy\rangle_{nd}=\langle\,x,y\rangle_{n}.
$$
By equation~~\eqref{adjoint_eq}, this is equivalent to
$V^*V=\id_n$, i.e., equation~~\eqref{conjug_eq}.
\end{proof}
\end{lem}

Let $\iso(\F,n,d)$ denote the set of matrices $V\in\M(\F^d,n)$
satisfying the equivalent
equations~~\eqref{isom_eq},~~\eqref{conjug_eq}. For
$V=[v_i^j]\in\iso(\F,n,d)$ we define the real $n\times n$ matrix
$P$ by
\begin{equation}      \label{map_eq}
p_i^j=\langle v_i^j, v_i^j\rangle=||v_i^j||^2.
\end{equation}
By equations~~\eqref{isom_eq} and~~\eqref{conjug_eq}, $P\in\B_n$.
Thus, equation~~\eqref{map_eq} defines a mapping
$$
\nu:\iso(\F,n,d)\to\B_n.
$$
\begin{defin}   \label{orthost_def}
A matrix $P\in\B_n$ is {\em $(\F,n,d)$-bistochastic}
($\F^d$-bistochastic for brevity) if $P=\nu(V)$ for some
$V\in\iso(\F,n,d)$.
\end{defin}

When $d=1$ and $\F=\R,\C,\HH$ Definition~~\ref{orthost_def} yields
orthostochastic, unistochastic, and qustochastic $n\times n$
matrices respectively. We will denote by $\os(\F,n,d)\subset\B_n$
the set of $\F^d$-bistochastic matrices.

\begin{prop}  \label{basic_prop}
\noindent{\em 1.} The set $\os(\F,n,d)\subset\B_n$ is closed.

\noindent{\em 2.} There are inclusions
$$
\os(\R,n,d)\subset\os(\C,n,d)\subset\os(\HH,n,d),
\os(\F,n,d)\subset\os(\F,n,d+1).
$$
\noindent{\em 3.} We have
$$
\os(\F,n,n)=\B_n.
$$
\begin{proof}
The set $\iso(\F,n,d)$ is a compact manifold and
$\nu:\iso(\F,n,d)\to\B_n$ is a differentiable map, yielding claim
1. Claim 2 is immediate from the definitions. We will prove claim
3. Let $P=[p_i^j]\in\B_n$ and let $d\in\N$ be arbitrary. Then
$P=\nu(V),V\in\iso(\F,n,d)$ if there are vectors $v_i^j\in\F^d$
such that $||v_i^j||^2=p_i^j$ and the vectors
$$
w_i=(v_i^1,\ldots,v_i^n)\in\F^{nd},\,1\le i \le n,
$$
are pairwise orthogonal. We have
\begin{equation}      \label{scalar_prod_eq}
\langle w_i,w_j\rangle_{nd}=\langle
v_i^1,v_j^1\rangle_{n}+\cdots+\langle v_i^n,v_j^n\rangle_{n}.
\end{equation}
If for $1\le k\le n$ the vectors $v_1^k,v_2^k,\ldots,v_n^k$ are
pairwise orthogonal in $\F^d$, then, by
equation~~\eqref{scalar_prod_eq}, the vectors $w_1,\ldots,w_n$ are
pairwise orthogonal in $\F^{nd}$. When $d\ge n$, the space $\F^d$
contains $n$ pairwise orthogonal vectors with arbitrary norms.
\end{proof}
\end{prop}

The following is immediate from Proposition~~\ref{basic_prop}.

\begin{cor}    \label{minimal_cor}
Let $\F$  be any of $\R,\C,\HH$. Then the following holds.

\medskip

\noindent{\em 1.} For any $n\in\N$ there is a unique
$d_{\min}(\F,n)\in\N$ such that $\os(\F,n,d_{\min})=\B_n$ and
$\os(\F,n,d)\ne\B_n$ for $d<d_{\min}$.

\noindent{\em 2.} We have
$$
d_{\min}(\HH,n)\le d_{\min}(\C,n)\le
d_{\min}(\R,n),d_{\min}(\F,n)\le d_{\min}(\F,n+1),
$$
and $d_{\min}(\F,n)\le n$.
\end{cor}

Let now $P\in\B_n$. By Proposition~~\ref{basic_prop}, there exist
$d\le n$ such that $P\in\os(\F,n,d)$. Let $d_{\min}(P,\F)$ be the
minimal such $d$. Then
\begin{equation}      \label{d_min_eq}
d_{\min}(\F,n)=\max\{ d_{\min}(P,\F):P\in\B_n\}.
\end{equation}
We will informally refer to $d_{\min}(P,\F),d_{\min}(\F,n)$) as
the {\em minimal number of internal degrees of freedom for
$\F$-quantization}.

\medskip

There are obvious identifications:
$$
\iso(\R,n,1)=O(n),\,\iso(\C,n,1)=U(n),\,\iso(\HH,n,1)=Sp(n).
$$
and
$$
\os(\R,n,1)=\oo_n,\,\os(\C,n,1)=\u_n,\,\os(\HH,n,1)=\hh_n.
$$
The following is well known:
$$
\os(\R,2,1)=\os(\C,2,1)=\os(\HH,2,1)=\B_2,
$$
$$
\os(\R,3,1)\subset\os(\C,3,1)=\os(\HH,3,1)\subset\B_3
$$
and the inclusions are proper. For $n>3$ there are proper
inclusions~~\cite{ChDo08}
$$
\os(\R,n,1)\subset\os(\C,n,1)\subset\os(\HH,n,1)\subset\B_n.
$$
\section{Relationships, symmetries, and dimension count}  \label{general}
There are several relationships between $(\F,n,d)$-orthostochastic
matrices for various values of $\F$ and $d$.

\begin{prop}  \label{inclusion_prop}
For any $n\ge 1$ and $d\in\N$ there are natural inclusions
$$
\os(\C,n,d)\subset\os(\R,n,2d),\ \os(\HH,n,d)\subset\os(\C,n,2d)
$$
and
$$
\os(\HH,n,d)\subset\os(\R,n,4d),
$$
\begin{proof}
The decomposition $z=x+\sqrt{-1}y$ identifies $\C^d$ and
$\R^{2d}$. Let $\langle\,,\rangle_{\C}$ and
$\langle\,,\rangle_{\R}$ be the complex and the real scalar
product on $\C^d$ respectively. The relationship
$$
\Re\left(\langle u,v\rangle_{\C}\right)=\langle u,v\rangle_{\R}
$$
yields the proper inclusion $\iso(\C,n,d)\subset\iso(\R,n,2d)$. It
is compatible with the {\em squared norm maps}
$\nu:\iso(\R,n,2d)\to\B_n,\,\nu:\iso(\C,n,d)\to\B_n$, yielding the
first inclusion. The second follows similarly from the isomorphism
$\HH^d=\C^{2d}$, and the third inclusion is the composition of the
former two.
\end{proof}
\end{prop}

Let $\W_n\subset\M(\R,n)$ be the group of permutation matrices. It
acts on $\M(\R,n)$ by left and by right multiplication, yielding
the action of $\W_n\times\W_n$ on $\M(\R,n)$ which preserves
$\B_n$. Let $\iso(\F^k)$ be the group of linear isometries of
$\F^k$. The set $\iso(\F,n,d)$ is invariant under precomposition
(resp. postcomposition) with elements in $\iso(\F^n)$ (resp.
$\iso(\F^{nd})$) yielding the action of
$\iso(\F^{nd})\times\iso(\F^n)$ on $\iso(\F,n,d)$. Let
$\diag(\F^n)\subset\iso(\F^n)$ (resp.
$\diag(\F,n,d)\subset\iso(\F^{nd})$) be the subgroup of diagonal
(resp. block-diagonal) isometries. By restriction, the subgroup
$\diag(\F,n,d)\times\diag(\F^n)$ acts on $\iso(\F,n,d)$. The group
$\W_n$ embeds naturally into $\iso(\F^n)$ (resp. $\iso(\F^{nd})$),
by permutation (resp. block permutation) matrices yielding an
action of $\W_n\times\W_n$ on  $\iso(\F,n,d)$.

\begin{prop}  \label{symmetry_prop}
The squared norm map $\nu:\iso(\F,n,d)\to\os(\F,n,d)$ is invariant
under the action of $\diag(\F,n,d)\times\diag(\F^n)$ and
equivariant for the actions of $\W_n\times\W_n$ on $\iso(\F,n,d)$
and $\os(\F,n,d)$.
\begin{proof}
Let $V=[v_i^j]\in\iso(\F,n,d)$. The action of $\diag(\F^n)$
multiplies the $n$ vectors $v_i^j\in\F^d$, where $i$ is fixed and
$1\le j\le n$, on the right by the same element $\la_i\in\F$ with
$||\la_i||=1$. The action of $\diag(\F,n,d)$ multiplies the $n$
vectors $v_i^j\in\F^d$, with $j$ fixed and $1\le i\le n$,  on the
left by the same isometry $U_j\in\iso(\F^d)$. These actions do not
change the norms of vectors $v_i^j$.

If $P=\nu(V)$ then $p_i^j=||v_i^j||^2$. The action of
$\W_n\times\W_n$ on $\iso(\F,n,d)$ permutes the rows and the
columns of $V$ the same way as its action on  $\B_n$ permutes the
rows and the columns of $P$.
\end{proof}
\end{prop}

By the preceding discussion, the group
$\diag(\F,n,d)\times\diag(\F^n)$ naturally acts on $\iso(\F,n,d)$.
We denote by $\doc(\F,n,d)$ the quotient space, i.e.,
$\doc(\F,n,d)=\diag(\F,n,d)\backslash\iso(\F,n,d)/\diag(\F^n)$. By
Proposition~~\ref{symmetry_prop}, the squared norm map
$\nu:\iso(\F,n,d)\to\B_n$ uniquely descends to a mapping of
$\doc(\F,n,d)$ which we will also denote by $\nu$. Thus,
$\os(\F,n,d)=\nu\left(\doc(\F,n,d)\right)\subset\B_n$. Let $\dim
X$ denote the real dimension.

\begin{prop}  \label{dimension_count}
The following equations hold:
\begin{equation}      \label{real_eq}
\dim\left(\doc(\R,n,d)\right)=(d-\frac12)n^2-\frac{d^2-d+1}{2}n,
\end{equation}
\begin{equation}      \label{complex_eq}
\dim\left(\doc(\C,n,d)\right)=(2d-1)n^2-(d^2+1)n+1,
\end{equation}
and
\begin{equation}      \label{quatern_eq}
\dim\left(\doc(\HH,n,d)\right)=(4d-2)n^2-(d^2+d+2)n.
\end{equation}
\begin{proof}
We have
$$
\dim\left(\iso(\R,n,d)\right)=(d-\frac12)n^2-\frac12n,\dim\left(\iso(\C,n,d)\right)=(2d-1)n^2,
$$
$$
\dim\left(\iso(\HH,n,d)\right)=(4d-2)n^2+n.
$$
Specializing to $d=1$, we recover the well known formulas
$$
\dim O(k)=\frac{k(k-1)}{2},\dim U(k)=k^2,\dim Sp(k)=2k^2+k.
$$
The groups $\diag(\F^n)$ satisfy
$$
\diag(\R^n)=\{\pm1\}^n,\diag(\C^n)=U(1)^n,\diag(\HH^n)=Sp(1)^n,
$$
The actions of $\diag(\F,n,d)$ and $\diag(\F^n)$ on $\iso(\F,n,d)$
are free and commute. They are transversal, except that both
$\diag(\C,n,d)$ and $\diag(\C^n)$ contain $U(1)$ as the group of
scalar unitary matrices. This information and the above formulas
yield the claims. We leave details to the reader.
\end{proof}
\end{prop}
%

%
%
%
%
%
%
%
%
%
%
%
%
%
%
%
%

%
\section{On $(n-1)$-orthostochasticity of $n\times n$ matrices}  \label{vector_ortho}
Let $P\in\B_n$. By Proposition~~\ref{basic_prop},
$d_{\min}(P,\R)\le n$. We will show that if $P$ satisfies mild
non-degeneracy assumptions and $n$ is odd, then $d_{\min}(P,\R)\le
n-1$. Let $n\in\N$. In what follows we will use the cyclic
convention for indices: $i+n=i$.

\begin{lem}           \label{linear_lem}
Let $n\in\N$ be an odd integer. Let $\xi_1,\dots,\xi_n$ be
arbitrary. Then the system
\begin{equation}   \label{linear_eq1}
x_i+x_{i+1}=\xi_{i+2},\,1\le i \le n,
\end{equation}
has a unique solution
\begin{equation}   \label{linear_eq2}
2x_i=-\xi_i+\xi_{i+1}+\xi_{i+2}-\xi_{i+3}+\xi_{i+4}-+\cdots+\xi_{i+n-1},\,1\le
i \le n.
\end{equation}
\begin{proof}
The matrix of the linear system~~\eqref{linear_eq1} is
nondegenerate, hence the solution is unique. The reader will
easily verify that equation~~\eqref{linear_eq2} yields a solution.
\end{proof}
\end{lem}
\begin{lem}           \label{skew_lem}
Let $n\in\N$ be an odd integer. Let $\xi_1,\dots,\xi_n$ be
positive numbers satisfying the inequalities
$$
\xi_i+\xi_{i+3}+\xi_{i+5}+\cdots+\xi_{i+n-2}\le\xi_{i+1}+\xi_{i+2}+\xi_{i+4}+\cdots+\xi_{i+n-1}
$$
for $1\le i \le n$. Then there exists a real skew-symmetric matrix
$A=[a_i^j]$ satisfying
\begin{equation}   \label{squares_eq}
\sum_{1\le i \le n}(a_i^j)^2=\xi_j,\ 1\le j \le n.
\end{equation}
\begin{proof}
Set $b_i^j=(a_i^j)^2$. Then $B=[b_i^j]$ is a symmetric $n\times n$
matrix with non-negative entries. The matrix $A$ satisfies
equation~~\eqref{squares_eq} if and only if $B$ satisfies
\begin{equation}   \label{sum_eq}
\sum_{1\le i \le n}b_i^j=\xi_j,\ 1\le j \le n.
\end{equation}

For $x_1,\dots,x_n\in\R_+$ set
$$
b_2^1=b_1^2=x_n,b_3^2=b_2^3=x_1,b_4^3=b_3^4=x_2,\dots,b_n^{n-1}=b_{n-1}^n=x_{n-2},
$$
$b_n^1=b_1^n=x_{n-1}$, and let $b_i^j=0$ for all other pairs of
indices. Then $B$ satisfies equation~~\eqref{sum_eq} if and only
if $x_1,\dots,x_n$ satisfy equation~~\eqref{linear_eq1}. The claim
now follows from Lemma~~\ref{linear_lem}.
\end{proof}
\end{lem}
\begin{thm}    \label{d_min_thm}
Let $n>1$ be an odd integer. Let $P=[p_i^j]$ be a $n\times n$
orthostochastic matrix. Suppose that $P$ satisfies the
inequalities $p_i^j\ne 0$ for $i\ne j$ and that for $1\le i \le n$
we have\footnote{We use the cyclic convention for indices.}
\begin{equation}   \label{inequal_eq2}
p_i^i+p_{i+3}^{i+3}+p_{i+5}^{i+5}+\cdots+p_{i+n-2}^{i+n-2}\le p
_{i+1}^{i+1}+p_{i+2}^{i+2}+p_{i+4}^{i+4}+\cdots+p_{i+n-1}^{i+n-1}.
\end{equation}
Then $P$ is $(n-1)$-orthostochastic.
\begin{proof}
We will find $V=[v_i^j]\in\iso(\R,n,n-1)$ such that $P=\nu(V)$.
For $1\le j \le n$ let
$v_1^j,\dots,v_{j-1}^j,v_{j+1}^j,\dots,v_n^j\in\R^{n-1}$ be
mutually orthogonal vectors such that $||p_k^j||^2=p_k^j$. The
vectors $\tv_k^j=v_k^j(p_k^j)^{-1/2}$ for $1\le k \le n,k\ne j,$
form an orthonormal basis in $\R^{n-1}$.

For $1\le i \le n$ let $w_i\in\R^{n(n-1)}$ and the notation
$\langle\,,\rangle_{n(n-1)}$ be as in the proof of
Proposition~~\ref{basic_prop}. For $1\le j \le n$ we set
\begin{equation}   \label{decompose_eq}
v_j^j=\sum_{k\ne j}a_k^j\tv_k^j.
\end{equation}
Thus, $w_1,\dots,w_n$ are the column vectors of $V=[v_i^j]$. The
equation $\nu(V)=P$ is equivalent to the two systems of quadratic
equations:
\begin{equation}   \label{quadratic_eq1}
\langle w_i,w_j\rangle_{n(n-1)}=0
\end{equation}
for $i\ne j$ and
\begin{equation}   \label{quadratic_eq2}
||v_j^j||^2=p_j^j
\end{equation}
for $1\le j \le n$.

By equation~~\eqref{decompose_eq}, we have
\begin{equation}   \label{skew_eq}
\langle w_i,w_j\rangle_{n(n-1)}=a_i^j+a_j^i
\end{equation}
for $i\ne j$ and
\begin{equation}   \label{norm_eq}
||v_j^j||^2=\sum_{i=1}^n(a_i^j)^2.
\end{equation}

Thus, it suffices to find a skew-symmetric $n\times n$ matrix
$A=[a_i^j]$ satisfying
\begin{equation}   \label{quadratic_eq3}
\sum_{i=1}^n(a_i^j)^2=p_j^j.
\end{equation}
%
%
Lemma~~\ref{skew_lem} yields a solution.
\end{proof}
\end{thm}
\begin{rem}     \label{explicit_rem}
{\em

The proof of Theorem~~\ref{d_min_thm} yields an explicit solution
of the equation $\nu(V)=P$. Let $e_1,\dots,e_{n-1}\in\R^{n-1}$ be
the standard orthonormal basis. For $1\le j \le n$ set
$v_k^j=(p_k^j)^{1/2}e_k$ if $k<j$ and $v_k^j=(p_k^j)^{1/2}e_{k-1}$
if $k>j$. The numbers $a_i^j$ defining the vectors
$v_j^j\in\R^{n-1}$ satisfy $a_i^j=0$ if $i-j\ne \pm 1\mod n$. For
pairs $i,j$ satisfying $i-j = \pm 1\mod n$,
Lemmas~~\ref{linear_lem},~~\ref{skew_lem} explicitly yield
$(a_i^j)^2$ as linear combinations of $p_k^k$ where $1 \le k \le
n$. Set $a_i^j=\sqrt{(a_i^j)^2}$ if $j<i$ and
$a_i^j=-\sqrt{(a_i^j)^2}$ if $i<j$.

}
\end{rem}
\begin{cor}    \label{open_set_cor}
Let $n>1$  be an odd integer. Then

\noindent{\em 1.} The set $\os(\R,n,n-1)$ contains all
bistochastic matrices $P=[p_i^j]$ such the numbers
$p_1^1,\dots,p_n^n$ satisfy the inequalities in
equation~~\eqref{inequal_eq2};

\noindent{\em 2.} For an open set of $P\in\B_n$ we have
$d_{\min}(P,\R)\le n-1$.
\begin{proof}
Let $P=[p_i^j]$ be a $n\times n$ bistochastic matrix. By
Theorem~~\ref{d_min_thm}, $P\in\os(\R,n,n-1)$ if the numbers
$p_1^1,\dots,p_n^n$ satisfy the strict inequalities in
equation~~\eqref{inequal_eq2} and $p_i^j\ne 0$ for $i\ne j$. This
is a nonempty open set, hence claim 2. By
Proposition~~\ref{basic_prop}, $\os(\R,n,n-1)$ contains its
closure, yielding claim 1.
\end{proof}
\end{cor}
\begin{exa}     \label{3_by_3_exa}
{\em

Let $n=3$. The open set in Corollary~~\ref{open_set_cor} is the
set of $P\in\B_3$ such that $p_i^j\ne 0$ for $i\ne j$ and
\begin{equation}   \label{3_by_3_eq}
p_1^1<p_2^2+p_3^3,\,p_2^2<p_3^3+p_1^1,\,p_3^3<p_1^1+p_2^2.
\end{equation}
Every $3\times 3$ matrix satisfying these inequalities is
$2$-orthostochastic. Note that the set of orthostochastic $3\times
3$ matrices has positive codimension in $\B_3$
\cite{AuPo79,ChDo08,Na96}. Inequalities~~\eqref{3_by_3_eq} hold if
and only if $p_1^1,p_2^2,p_3^3$ are the side lengths of a
nondegenerate triangle. We point out that the triangle
inequalities come up as conditions of unistochasticity for
$3\times 3$ matrices \cite{DuZy11}.

}
\end{exa}
\begin{exa}     \label{5_by_5_exa}
{\em

Let $n=5$. The set of $4$-orthostochastic $5\times 5$ matrices in
Corollary~~\ref{open_set_cor} consists of bistochastic matrices
satisfying the following:
$$
p_1^1+p_4^4 \le p_2^2+p_3^3+p_5^5,\,p_2^2+p_5^5 \le
p_3^3+p_4^4+p_1^1,
$$
and
$$
p_3^3+p_1^1 \le p_4^4+p_5^5+p_2^2,\,p_4^4+p_2^2 \le
p_5^5+p_1^1+p_3^3,\,p_5^5+p_3^3 \le p_1^1+p_2^2+p_4^4.
$$
(These inequalities do not have an immediate geometric
interpretation.) The interior of the set of $4$-orthostochastic
$5\times 5$ matrices contains $[p_i^j]\in\B_5$ such that $p_i^j\ne
0$ for $i\ne j$ and the above inequalities are strict.

}
\end{exa}
\section{Concluding discussion}  \label{conj}
To describe precisely the sets of orthostochastic, unistochastic,
and qustochastic $n\times n$ matrices for arbitrary $n$ seems a
difficult problem {\bf ref}. The concept of $\F^d$-bistochastic
introduced here replaces this problem with another, seemingly
simpler, but still a meaningful problem: To characterize
$\F^d$-bistochastic $n\times n$ matrices for arbitrary $n$. Our
results suggest that this question is still nontrivial but less
subtle. The two questions are related. Note that, by
Proposition~~\ref{inclusion_prop}, the set of $2$-orthostochastic
matrices contains the set of unistochastic matrices, and the set
of $2$-unistochastic matrices contains the set of qustochastic
matrices. Let $n$ be arbitrary, and let us vary $d\in\N$. By
Proposition~~\ref{basic_prop}, the set of $\F^d$-bistochastic
coincides with $\B_n$ as $d$ reaches $d_{\min}(\F,n)\le n$. The
above results and dimensional considerations suggest that
$d_{\min}(\F,n)$ does not indefinitely increase, as $n$ goes to
$\infty$. In particular, for $\F=\R$, the following conjecture is
plausible.

\vspace{4mm}

\begin{conj}     \label{main_conj}
Denote by $d_{\min}(n)$ the smallest value of $d$ such that the
set of $d$-orthostochastic matrices coincides with $\B_n$. Then
there exists $n_0$ such that for $n\ge n_0$ we have
$$
2 \le d_{\min}(n) \le 3.
$$
\end{conj}

\medskip

The material in section~~\ref{vector_ortho} suggests that value of
$d_{\min}(n)$ may depend on the parity of $n$. The author believes
that the threshold dimension should not be too large. Most likely,
$n_0=3$.

\medskip

\noindent Acknowledgements. The work was partially supported by
the MNiSzW grant N N201 384834 and the NCN Grant
DEC-2011/03/B/ST1/00407.

\end {document}